\documentclass[aip,reprint,pof]{revtex4-1}

\usepackage{graphicx}
\usepackage{subfigure}
\usepackage{amsmath}
\usepackage{amssymb}
\usepackage{latexsym}
 \usepackage{cancel}
 \usepackage{multirow}
 \usepackage{color}
 \usepackage{epstopdf}

\newcommand{\ve}[1]{\mbox{\boldmath $#1$}}

\draft 

\begin{document}


\title{Rayleigh-Taylor instability under an inclined plane} 



\author{P.-T. Brun}
\email[]{ptbrun@mit.edu}
\affiliation{Laboratory of Fluid Mechanics and Instabilities, EPFL, 1015 Lausanne, Switzerland}
\affiliation{Department of Mathematics, Massachusetts Institute of Technology, Cambridge, Massachusetts 02139, USA}

\author{Adam Damiano}
\affiliation{Laboratory of Fluid Mechanics and Instabilities, EPFL, 1015 Lausanne, Switzerland}

\author{Pierre Rieu}
\affiliation{Laboratory of Fluid Mechanics and Instabilities, EPFL, 1015 Lausanne, Switzerland}

\author{Gioele Balestra}
\affiliation{Laboratory of Fluid Mechanics and Instabilities, EPFL, 1015 Lausanne, Switzerland}

\author{Fran\c{c}ois Gallaire}
\affiliation{Laboratory of Fluid Mechanics and Instabilities, EPFL, 1015 Lausanne, Switzerland}


\date{\today}

\begin{abstract}
We revisit the canonical Rayleigh-Taylor instability and investigate the case of a thin film of fluid upon the underside of an inclined plane. The presence of a natural flow along the plane competes with the conventional droplet forming instability. In particular, experiments reveal that no drops form for inclinations greater than a critical value. These features are rationalized in the context of the absolute/convective analysis conducted in this article. 
\end{abstract}


\maketitle 



%
%

%

\section{Introduction}
\begin{figure}
\begin{center}
\includegraphics[width=\textwidth]{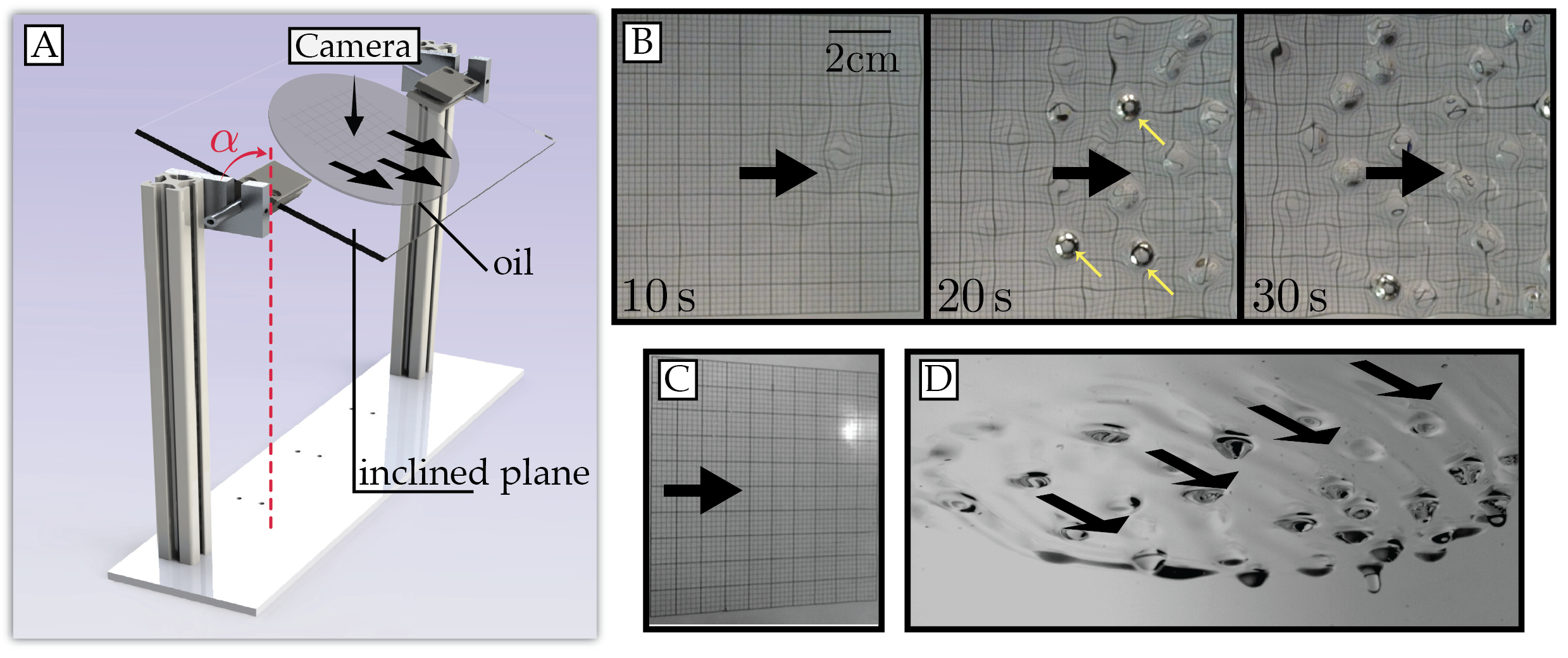}
\caption{ (a) A thin film of initial thickness $h_i$ flows on the underside of a transparent inclined plane. The angle $\alpha$ denotes the inclination of the substrate with respect to the gravity $g$. (b) A typical experimental observation for $h_i=0.52\,\ell_c$ and $\alpha=81^\text{o}$. (c) No drops are formed for $h_i=0.55\,\ell_c$ and $\alpha=60.5^\text{o}$. (d) For large enough values of $\alpha$, droplets form according to the Rayleigh-Taylor instability. The black arrows indicate the overall direction of the flow.   }
\label{fig1}
\end{center}
\end{figure}

The Rayleigh-Taylor instability (RTI) occurs when a fluid is momentarily sustained above a fluid of lesser density~\cite{Sharp:1984wk}. Examples are found in a variety of situations ranging from the simple case of water sitting on oil in a kitchen glassware to supernova explosions~\cite{wang2001instabilities}. This interfacial instability has been thoroughly studied and has served as a foundation for tackling fundamental issues such as the break up of free-surface flows~\cite{Eggers:2008hq} (drop formation) and pattern generation \cite{Fermigier:1992vm}.  Applicative components of this instability also arise due to its natural tendency to enhance the mixing of liquid species~\cite{Sharp:1984wk}. Additionally, the RTI is a prime concern when coating surfaces, be it with paint or a lubricant, as the instability may lead to coating irregularities or to the detachment of droplets for thick coatings. As such, many studies have concentrated on means of controlling or suppressing the growth of pendant drops when applying temperature gradients, electric fields, or surface tension gradients (see Ref.\ \cite{Weidner:2007vo} for example). In the following, we revisit the iconic problem of a thin liquid film spread upon the underside of a flat surface and investigate the effect of the substrate orientation on the instability.

In the canonical RTI of a thin viscous fluid layer coated on the underside of a \textit{horizontal} surface, the interface perturbations are found to grow exponentially in time and generate drops arranged in various patterns ~\cite{Fermigier:1992vm}. The instability development is set by the competition of surface tension effects and gravitational effects such that the distance between droplets is found to be $\lambda=2\pi\sqrt{2}\ell_c$ where $\ell_c=\sqrt{\gamma/\rho g}$ is the capillary length of the problem and $\gamma$, $\rho$, and $g$ denote the thin layer surface tension, its density, and the acceleration of gravity, respectively. This natural length scale is well recovered with linear arguments since $\lambda$ is the wavelength of the most unstable mode predicted by linear stability analysis.

Herein, we extend the study to cases where the surface is \textit{tilted}  and set at an angle $\alpha\leqslant\pi/2$ from the vertical (see Fig.~\ref{fig1}, $\alpha=\pi/2$ corresponds to the horizontal case). In particular, we focus on how the main features of the instability are affected by the base flow along the substrate. For values of $\alpha\simeq\pi/2$  we observe that the interface deforms and eventually forms droplets according to the RTI (Fig.~\ref{fig1}(b,d)). For values of $\alpha\lesssim\pi/2$, deformations of the surface are greatest in the downstream direction where the fluid accumulates; this is also the area where the first droplets form. Soon after, more droplets form and drip from the upward area as well. However, building upon the known characteristics of flows on vertical walls -- be it from the application of varnish on furniture in daily experience or else when referring to the literature on plates extracted from liquid baths~\cite{landaulevich} -- one may anticipate that the inclination of the substrate is likely to affect the instability to a point where it could disappear, say at a critical angle $\alpha_c$. This angle  $\alpha_c$ has been observed experimentally (see  Fig.~\ref{fig1}(c)) as explained in the results Sec.~\ref{res}.  We will see that traditional temporal linear stability arguments, which are sufficient in the horizontal case, fail to predict the value of $\alpha_c$. The experiments are instead  rationalized by addressing the absolute or convective nature of the aforementioned linear stability results as detailed hereafter. Note that the Kapitza instability~\cite{kapitza1949wave,kalliadasis2011falling} is here neglected owing to the low Reynolds numbers characterizing the experiment. Hence, we solely focus our analysis on the RTI.

\section{The flow under an inclined plane}
\label{linear}

\begin{figure}[!h]
\begin{center}
\includegraphics[width=.77\textwidth]{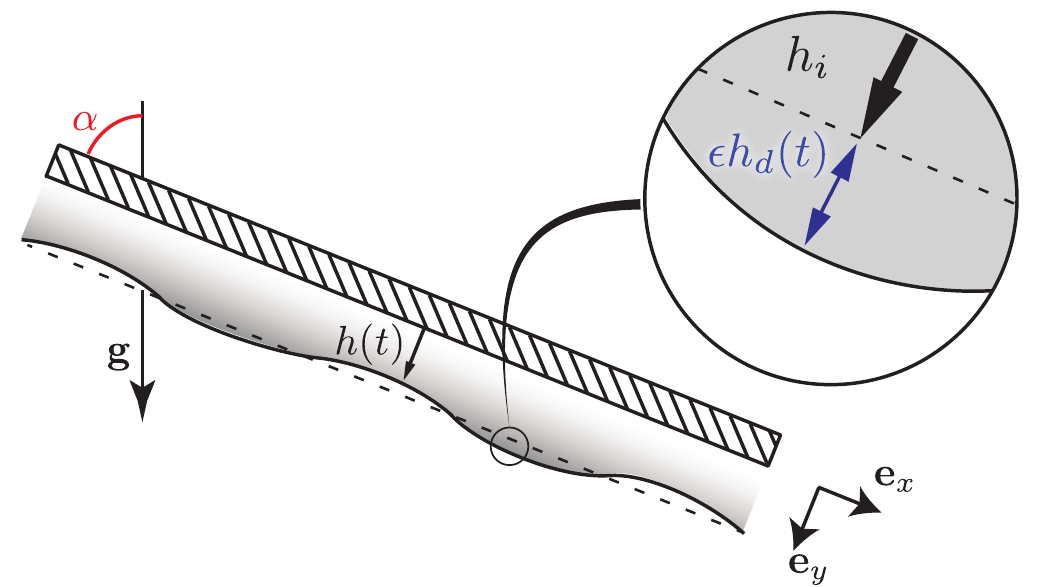}
\caption{A thin film of initial thickness $h_i$ flows under an inclined plane in the direction $\mathbf{e}_x$. The angle $\alpha$ denotes the inclination of the substrate with respect to the direction of gravity $\mathbf{g}$.  }
\label{fig2}
\end{center}
\end{figure}

We consider the flow of a thin fluid film of initial thickness $h_i$ placed underneath an inclined plane (Fig.~\ref{fig1}). The film, of density $\rho$, viscosity $\mu$, and surface tension $\gamma$ is subject to the acceleration of gravity $g$. Of particular importance herein is the angle $\alpha>0$ between the substrate and the vertical. 
The thickness $h_i$ is supposed smaller than the typical length scale over which the coating spreads, i.e., the problem is assumed invariant in the flow direction. Additionally, mass conservation indicates that the flow velocity $\ve u =(u,v)$ is essentially one dimensional  ($v\sim u \times h/L \ll u$).  
Assuming low Reynolds number, one may write the lubrication equations~\cite{leal2007advanced} for this flow yielding:
\begin{equation}
\label{eq:u2}
u (x,y) = \frac{\rho g}{\mu} \left( -\sin \alpha \frac{\partial h}{\partial x} - \frac{\gamma}{\rho g }\frac{\partial \kappa}{\partial x}- \cos \alpha\right)y(y-2h)
\end{equation}
where $\kappa \simeq \frac{\partial^2 h}{\partial x^2}$ is the interface curvature. 
Integrating eq.~(\ref{eq:u2}) leads to finding the flow rate $Q\propto h^3$, which combined to mass conservation yields:
\begin{equation}
\label{eq:mc}
\frac{\partial h}{\partial t} =-\frac{\partial }{\partial x} \left( \frac{\rho g h^3}{3 \mu}\left( \sin \alpha \frac{\partial h}{\partial x}  +\ell_c^2 \frac{\partial \kappa}{\partial x} \right)\right)-\frac{\rho g h^2}{ \mu} \frac{\partial h}{\partial x}\cos \alpha \, .
\end{equation}
The trivial solution $h=h_i$  satisfies  eq.~(\ref{eq:mc}) and is valid for any value of $h_i\ll L$. In that case, the Poiseuille like flow in the $x$-direction is such that the free surface velocity writes:

\begin{equation}
\label{eq:ui}
	u_i=\cos{\alpha}\frac{\rho g}{\mu}h_i^2 \, .
\end{equation} 

We now turn to examining the stability of the uniform solution to linear perturbations. Assuming a normal decomposition $h(x,t)=h_i+\epsilon h_d(x,t)$ with $\epsilon \ll 1$ and $h_d(x,t)\propto e^{i(kx-\omega t)} $ (see Fig.~\ref{fig2}), one obtains the following dispersion relation for $\omega$ from eq.~(\ref{eq:mc}):

\begin{equation}
\label{dispersion}
\omega (k)=  \frac{\rho g h_i^3 }{3 \mu} \left( i\left(  k^2 \sin \alpha- \ell_c^2 k^4\right)\right)
+ \left(\frac{\rho g h_i^2}{\mu} \cos \alpha \right)k \,.
\end{equation}
The real part of $\omega$ represents the advection of the wave and is found to be equal to $u_i$, the velocity at the interface of a uniform film. The imaginary part of $\omega$, denoted $\sigma$, is the temporal growth rate of the instability, whose sign sets the stability of the perturbation. The flow is either stable, marginally stable, or unstable for $\sigma<0$, $\sigma=0$, and $\sigma>0$ respectively. 

\begin{figure}[!h]
\begin{center}
\includegraphics[width=\textwidth]{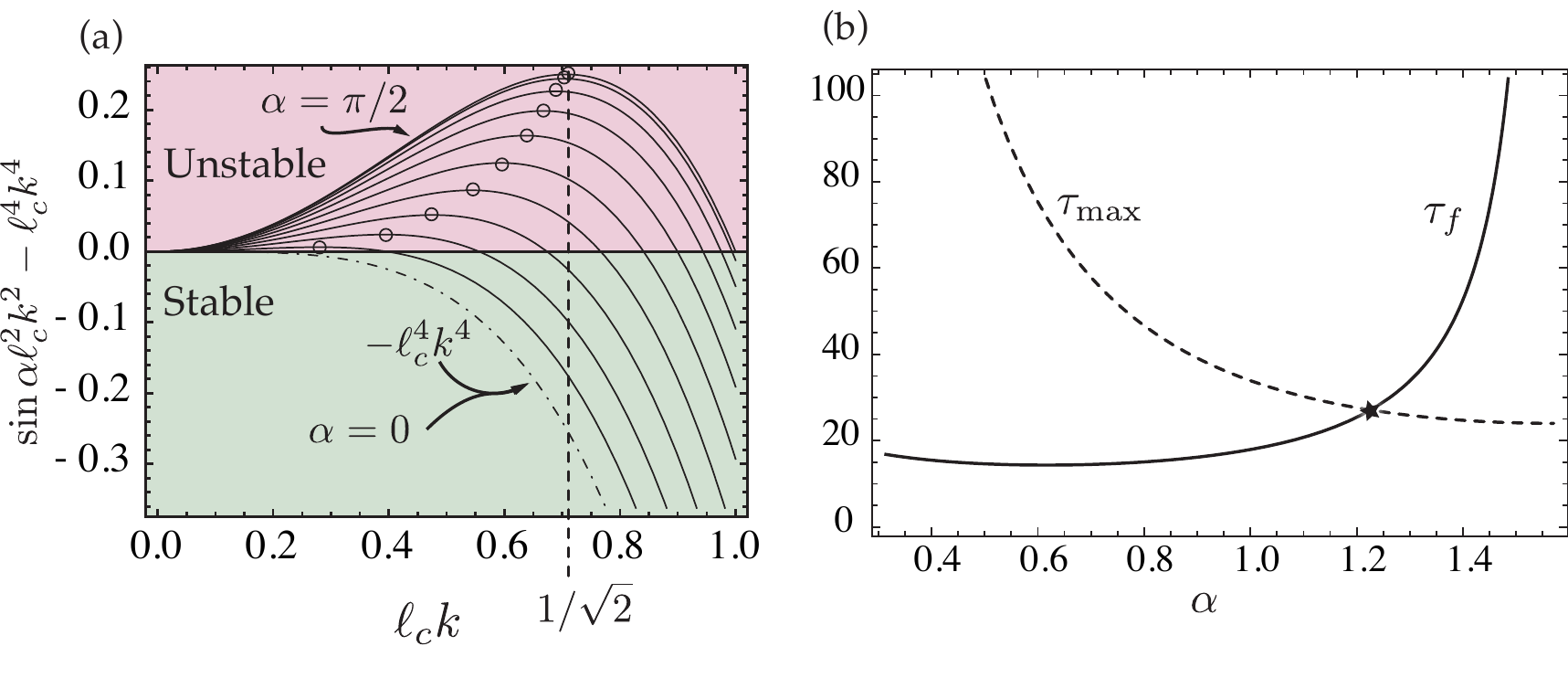}
\caption{(a) Growth rates for inclinations varying from $\pi/2$ to 0 with a $\pi/20$ increment. The most unstable mode is marked by a black circle. All plain lines denotes unstable cases and the dotted dash line (corresponding to $\alpha=0$) is found marginally stable. For $\alpha=\pi/2$ the maximum growth rate corresponds to $k_\textrm{max}=\ell_c^{-1}/\sqrt{2}$. (b) Characteristic times of the instability growth ($\tau_\textrm{max}$, plain line) and of its advection ($\tau_f$). The dash line corresponds to $h_i/\ell_c=0.5$.  }
\label{growth}
\end{center}
\end{figure}

As one may observe in Fig.~\ref{growth}, the flow is always found to be linearly temporally unstable for strictly positive values of $\alpha$ and the most unstable modes are: 
\begin{equation}
k_{\textrm{max}}=\frac{1}{\ell_c\sqrt{2}}\sqrt{\sin{\alpha}}.
\label{max}
\end{equation}
They correspond to a wave length $\lambda_{\textrm{max}}=2\pi\sqrt{2}\ell_c/\sqrt{\sin\alpha}$ and comments may be made regarding this expression. First, when substituting $\alpha=\pi/2$ in eq.~(\ref{max}) one recovers the wavelength of the RTI of a thin film underneath an horizontal plate, $\lambda_{\alpha=\pi/2}=2\pi\sqrt{2}\ell_c$, which sets the distance between neighboring droplets (on the order of 2cm here). Second, the sensitivity of $\lambda_{\textrm{max}}$ with respect to $\alpha$ is limited when $\alpha\simeq\pi/2$ (see the $\sqrt{\sin{\alpha}}$ in eq.(\ref{max}) and the proximity of the higher most two curves in Fig.~\ref{growth}(a)). Finally, a crude interpretation of  eqs.~(\ref{dispersion}) and (\ref{max}) would lead to the conclusion that  only perfectly vertical walls prevent drops from forming. In Fig.~\ref{growth}(a) the dot-dash curve corresponding to $\alpha=0$ is the only one fully enclosed in the stable region. In other words, the value of the earlier defined critical angle would be $\alpha_c=0$. This is in contradiction with our experimental observation presented in Fig.~\ref{fig1}(c) where $\alpha\simeq\pi/3$ and no drops form. This paradox will be solved by the absolute/convective analysis in Sec.~\ref{absconvsection}, but a first insight in the mechanism at stake may be gained when evaluating the typical time scales of the problem.  

We now proceed to evaluate and compare the characteristic times of the instability and of the flow. Let us define $\tau_\text{max}$ to be the inverse of the growth rate of the most unstable mode, i.e., $\tau_\textrm{max}=\sigma(k_\textrm{max})^{-1}$.  Using eqs.~(\ref{dispersion}) and (\ref{max}) we find that: 
\begin{equation}
\label{max2}
\tau_\textrm{max}\propto \frac{1}{\sin^2\alpha}
\end{equation}
Similarly, $\tau_f$ is defined as the time of advection over the wavelength $\lambda_\textrm{max}$, i.e., $\tau_f\simeq\lambda_\textrm{max}/u_i$ yielding: 
\begin{equation}
\label{ }
\tau_f\propto\frac{1}{\cos{\alpha\sqrt{\sin\alpha}}}
\end{equation}

As shown in Fig.~\ref{growth}(b), $\tau_f$ and $\tau_\textrm{max}$ are respectively increasing and decreasing with $\alpha$. In particular, they intersect for a given angle denoted $\alpha^*$.  The practical implications are as such: for small angles $\alpha<\alpha^*$ we anticipate that the instability will be dominated by the flow since the time required for the instability to form ($\tau_\textrm{max}$) is much larger than the typical time needed for the same instability to flow down $\tau_f$. Conversely, at large angles the instability develops much faster than the fluid flows, i.e., we anticipate that drops will form as if there were no flow (in the limit of very large angles $\alpha\simeq\pi/2$). The distinction between these two regimes may be made rigorously in the context of a spatio-temporal analysis as detailed in the following. 

\section{Absolute and convective instabilities}
\label{absconvsection}

In this section, we make use of the absolute/convective instability concepts to analyze the RTI on the underside of an inclined surface.
The distinction between absolute and convective instability has been pioneered in plasma instabilities \cite{briggs1964electron,Bers1983} and then more recently applied to fluid-dynamical instabilities in parallel and weakly spatially developing open flows \cite{huerre1990local}. The nature of a given flow depends on the large-time asymptotic behavior of the linear impulse response; the flow is convectively unstable if the amplified disturbances move away from the source, conversely, the flow is absolutely unstable when amplified perturbations invade the entire flow. In contrast to the temporal stability problem where the axial wavenumber $k$ is real and one seeks a complex frequency $\omega$, the absolute/convective nature of the instability is determined by applying the Briggs-Bers zero-group velocity criterion to the dispersion relation for fully complex $(k,\omega)$ pairs~\cite{briggs1964electron,Bers1983}. In order to determine the transition from convective to absolute instability, it is sufficient to detect the occurrence of saddle points in the characteristics of spatio-temporal instability waves, i.e., a complex value of the wavenumber $k_0$ such that $\frac{\partial  \omega_r}{\partial  k_r}=\frac{\partial  \omega_i}{\partial  k_r}=0$.
The discrimination between absolutely and convectively unstable flows was shown to play a crucial role in accounting for the occurrence of synchronized self-sustained oscillations in a variety of spatially developing shear flows, such as single phase wakes, hot jets, and counter-flow mixing layers \cite{huerre1990local}. But it also explains, for instance, the transition from dripping to jetting in two-phase immiscible microfluidic co-axial injectors \cite{guillot2007stability}, as well as different regimes occurring in the pearl forming instability of a film flowing down a fiber \cite{duprat2009spatial}.

In anticipation of the following analysis we rewrite eq.~(\ref{dispersion}) in a convenient dimensionless form:
\begin{equation}
\label{dispersionadim}
\tilde{\omega} =  i\left(\tilde{k}^2- \tilde{k}^4\right)+ \tilde{u} \tilde{k} \, ,
\end{equation}
where the wavenumber and angular frequency have been non-dimensionalised with the characteristic space and time scales as:
\begin{equation}
\label{adim}
\tilde{k} = k \frac{\ell_c}{\sqrt{\sin \alpha}} \,\,\,  \textrm{and}\,\,\, \tilde{\omega}=\omega  \frac{3 \ell_c^2\mu}{\rho g h_i^3 \sin^2 \alpha }.
\end{equation}
The non-dimensional interface velocity reads:
\begin{equation}
\label{Uadim}
\tilde{u}= \frac{3}{\sqrt{\sin \alpha}\tan \alpha}\frac{\ell_c}{h_i} \, .
\end{equation}

Hereafter, both the wavenumber $\tilde{k}$ and the wave-function $\tilde{\omega}$  are complex numbers. Denoting $\tilde{k}=\tilde{k}_r+i\tilde{k}_i$ and injecting this form in eq.~(\ref{dispersionadim}) yields:
\begin{align}
\tilde{\omega}_r &= \tilde{k}_r \left(-4 \tilde{k}_i ^3+\tilde{k}_{i} \left(4 {\tilde{k}_r}^2-2\right)+\tilde{u}\right)   \\
\label{omi}
\tilde{\omega}_i &=  -{\tilde{k}_i}^4+{\tilde{k}_i}^2 \left(6 {\tilde{k}_r}^2-1\right)+{\tilde{k}_i} \tilde{u}-{\tilde{k}_r}^4+{\tilde{k}_r}^2
\end{align}
The complex group velocity writes $ \tilde{v}_g=\frac{\partial \tilde{\omega}}{\partial \tilde{k}}$ and we denote $\tilde{\omega}_0$ and $\tilde{k}_0$ the absolute frequency and wavenumber, respectively. They are defined by the zero-group-velocity condition $\tilde{v}_g(\tilde{k}_0)=0$, such that $\tilde{\omega}_0=\tilde{\omega}(\tilde{k}_0)$. 
Differentiating (\ref{omi}) with respect of $\tilde{k}_r$ and $\tilde{k}_i$ and canceling both equations, one obtains:
\begin{align}
\label{crit1}
\tilde{u}=-32\tilde{k}_{0,i}^3-4\tilde{k}_{0,i} \\
\label{crit2}
\tilde{\omega}_{0,i}= 1/4 - 2 \tilde{k}_{0,i}^{2} - 24 \tilde{k}_{0,i}^{4}
\end{align}
Note that the values of $\tilde{k}_{0,i}$ and $\tilde{\omega}_{0,i}$ are directly set by the value of the parameter $\tilde{u}$, itself function of the physical parameters $h_i/\ell_c$ and $\alpha$ (see eq.~(\ref{Uadim})). In turn, the value of $\tilde{\omega}_{0,i}$ relative to zero sets the nature of the flow.
When $\tilde{\omega}_{0,i}<0$ the flow is convectively unstable and conversely, it is absolutely unstable when  $\tilde{\omega}_{0,i}>0$. In both cases the flow is linearly unstable, as found earlier in Sec.~\ref{linear}. However, those two cases lead to completely different scenarios, which are the main focus of the rest of the article.  
We now search for the roots of eq.~(\ref{crit2}), which combined with eq.~(\ref{crit1}), yields to the exact expression of the critical value of $\tilde{u}=\tilde{u}^*$ leading to a transition between the convective and absolute regimes, namely: 
\begin{equation}
\label{Ustar}
\tilde{u}^*= \frac{1}{\sqrt{27}}\sqrt{34+14 \sqrt{7}} \simeq 1.62208 \, .
\end{equation}

\begin{figure}
\begin{center}
\includegraphics[width=\textwidth]{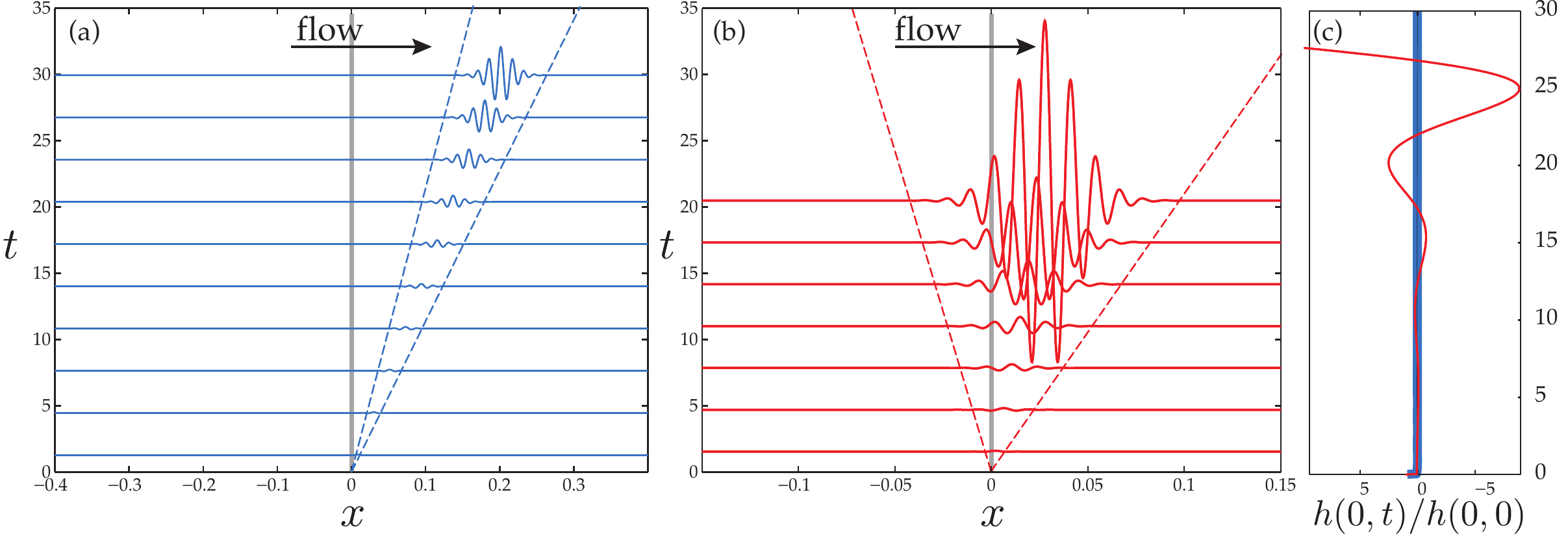}
\caption{Spatio-temporal evolution of two wave packets generated by a Dirac perturbation in $x=0$ at $t=0^+$ for $h_i/\ell_c\simeq 0.67$ and (a) $\alpha=\pi/4$ and (b)  $\alpha=\pi/2.2$. Both cases are unstable, but (a) is convectively unstable, the perturbation decreases with time in $x=0$, and conversely, (b) is absolutely unstable as the perturbation grows exponentially in the laboratory frame. Dashed lines correspond to the fronts of the perturbed wedges. (c) shows the time evolution of the relative amplitude of the perturbation at the origin in both cases.}
\label{gio}

\end{center}
\end{figure}

With a dispersion relation taking the form found in eq.~(\ref{dispersionadim}), the leading $\tilde{v}_f^{+}$ and receding $\tilde{v}_f^{-}$ front velocities of the wedge of the impulse response can be directly determined as the sum of the inherent advection $\tilde{u}$ and $\pm \, \tilde{u}^*$:
\begin{equation}
\label{eq:frontVelocities}
	\tilde{v}_f^{\pm} =  \tilde{u} \pm \tilde{u}^{*}   = \tilde{u} \pm \frac{\Delta \tilde v_f}{2} \, ,
\end{equation}
where $\Delta \tilde v_f = \tilde v_f^+ - \tilde v_f^- = 2\tilde{u}^{*}$ is the width of the perturbed region.
This concept is best illustrated when following the time evolution of a wave packet generated with a localized initial perturbation. In Fig.~\ref{gio}, we report the time evolution of two packets generated by a Dirac impulse in $x=0$ at $t=0$ and observed  in the laboratory frame. 
First, a global view of the flow as a function of time is presented in the two panels in Fig.~\ref{gio}(a) and (b).  They correspond to  $\tilde{\omega}_{0,i}<0$ and $\tilde{\omega}_{0,i}>0$, respectively. Second, the time evolution at the impulse location, $h(0,t)$, is presented in Fig.~\ref{gio}(c). 
When  $\tilde{\omega}_{0,i}<0$, the perturbation increases with time but is carried away faster than it spreads across the flow and the upstream front velocity $\tilde{v}_f^{-}$ of the perturbed wedge is positive. As a consequence, the perturbation at the source point ($x=0$) decreases with time such that $h(0,t)$ is quickly indistinguishable from the vertical axis, and the instability is convective. Conversely, if  $\tilde{\omega}_{0,i}>0$, the perturbation grows exponentially in the laboratory frame despite the presence of the flow. The upstream front velocity $\tilde{v}_f^{-}$ of the perturbed wedge is negative, so the wave-packet counter-propagates to invade the all space, and the instability is absolute (Figure~\ref{gio}b). 
These examples emphasize that the absolute/convective transition is found when $\tilde v_f^{-} =0$, i.e., $\tilde u = \tilde u^*$.

Recalling from eqs.~(\ref{Uadim}) and  (\ref{Ustar}) that $\tilde u$ is a decreasing function of $\alpha$ and that $\tilde u^*$ is constant, one may derive the critical value of the inclination $\alpha^*$ yielding the absolute/convective transition:
\begin{equation}
\label{criteria}
\sqrt{\sin \alpha^*}\tan \alpha^*=\frac{3\ell_c}{\tilde{u}^*h_i}\,.
\end{equation}
The existence of such an angle had been anticipated in the time scale analysis in Sec.~\ref{linear} using temporal arguments. However, the presented spatio-temporal analysis constitutes a rigorous derivation of its exact expression.  Note that these results are consistent with the mechanism depicted in Sec.~\ref{linear}. Representing eq.~(\ref{eq:frontVelocities}) with dimensional terms, the existence of a critical angle results from the competition of the advection velocity $u_i=\cos{\alpha}\frac{\rho g}{\mu}h_i^2$, which increases with an increasing inclination (decreasing values of $\alpha$), and the wedge spreading of the unstable region:
\begin{equation}
\label{eq:wedgeWidth}
	\Delta v_f = 2\tilde{u}^{*} \frac{\rho g h_i^3 \sin^{3/2} \alpha }{3 \ell_c \mu} \,,
\end{equation}
which decreases with an increasing inclination (decreasing values of $\alpha$). A critical angle is unavoidable as an increasing inclination of the plane therefore simultaneously increases the advection and decreases the growth rate.

We turn back to examining the critical angle in eq.~(\ref{criteria}). This implicit expression is solely a function of the ratio $h_i/\ell_c$, such that surface tension and gravity effects influence the problem through the capillary length $\ell_c$. On the other hand, the fluid viscosity $\mu$ does not enter in eq.~(\ref{criteria}). This could have been anticipated as $\mu$ intervenes linearly both in the typical flow rate and in the typical droplet formation time, such that it does not affect their relative values.
\begin{figure}[!h]
\begin{center}
\includegraphics[width=\textwidth]{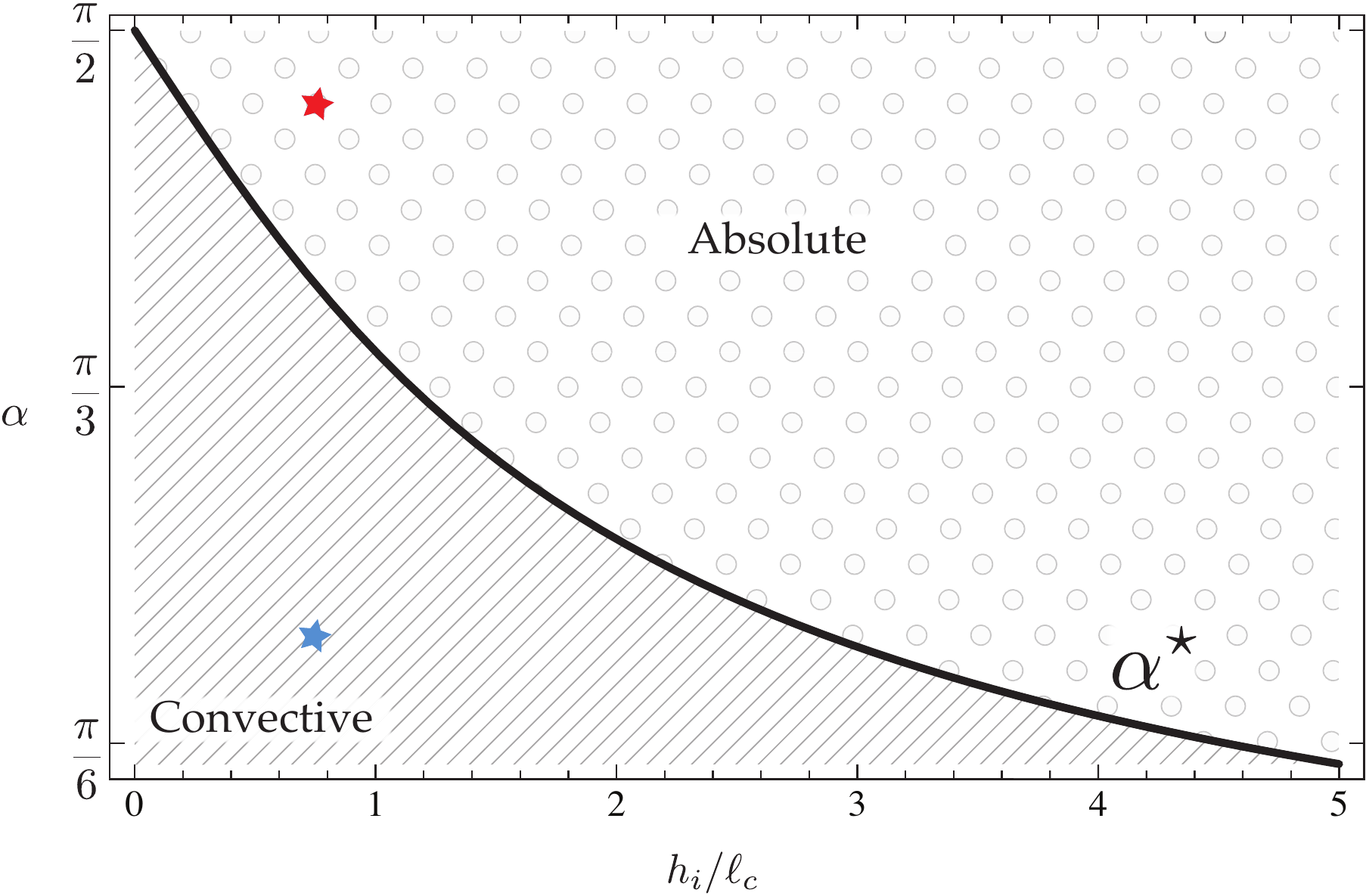}
\caption{State diagram of the flow under an inclined plane in the parameter space $(h_i/\ell_c,\alpha)$. Plotted in black are the values of the critical angle $\alpha^*$ delimiting the absolute and convective domains. The two stars indicate the physical parameters used in Fig.~\ref{gio}. }
\label{alphaplot}
\end{center}
\end{figure}
A practical consequence of eq.~(\ref{criteria}) is the state diagram shown in Fig.~\ref{alphaplot} where the plot of $\alpha^*$ as a function of $h_i/\ell_c$ may be found. For a given value of $h_i/\ell_c$, the flow is absolutely unstable for angles $\alpha>\alpha^*$ and convectively unstable for $\alpha<\alpha^*$. Note that for $\alpha=\pi/2$, all thin films with non-zero thickness are absolutely unstable since such an angle forbids any  flow parallel to the substrate to compete with the instability. 

In summary,  the flow under an inclined plane is always linearly temporally unstable (unless $\alpha=0$). However, two cases are possible. Depending on the respective values of $\alpha$ and $h_i/\ell_c$ the flow is either convectively or absolutely unstable. In the first case, perturbations are found to decrease in the laboratory frame, despite the unstable nature of the flow, as they are carried away by advection.  In the second, the instability is strong enough to overcome the flow. We now examine experimentally the consequences of such scenarios. 


\section{Experiments}
\label{res}
\begin{figure}
\begin{center}
\includegraphics[width=\textwidth]{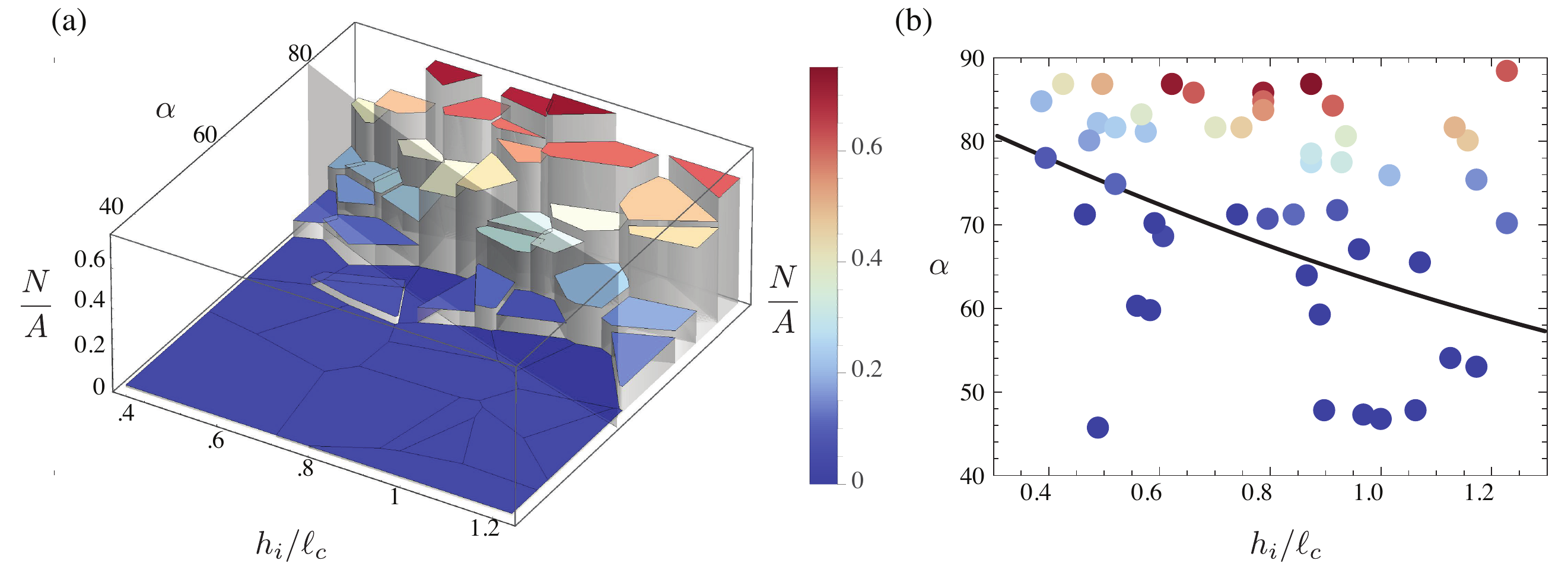}
\caption{(a) Experimental results. Each experiment is represented by the triplet ($h_i/\ell_c,\alpha, \bar N$) and is color coded according to the scaled number of droplets formed $\bar N$. (b) Experimental data compared to the theoretical expression $\alpha^*$. }
\label{exp}
\end{center}
\end{figure}

The experiments were conducted on a test stand which consists of a base plate, two columns, and two mounting plates where the clamp assemblies that hold the experimental surface in place are free to rotate (Fig.~\ref{fig1}(a)). The working fluid used in the experiments is Castor Oil  (H\"anseller AG).  The value of the capillary length $\ell_c=1.91$\text{\,mm}  was measured using Axisymmetric Drop Shape Analysis~\cite{del1997axisymmetric} on a pendant drop experiment. Plexiglas was chosen to serve as support surfaces for the experiments. A grid was included in the setup to enhance the distortions of the interface. The grid was aligned parallel with the surface and placed far below the flow so as not to induce any perturbations. 

A typical experiment is divided in two steps: preparation and testing. First, a fluid film is prepared on the  solid surface: the surface is made horizontal and a mass $m$ of oil is deposited in the center of the surface so that it will spread over time. Its area $A$ increases with time, $A\propto t^{1/8}$, while its thickness remains uniform apart from the boundaries~\cite{huppert1982propagation}. Second, at a given time $t=t_i$ the surface is gently rotated into the testing position where the fluid is beneath the surface, which makes an angle $\alpha$ with the vertical axis $\mathbf e_z$ (Fig.~\ref{fig1}(a)). At the beginning of the testing, the thickness of the film writes: $h_i=\frac{m}{\rho}\frac{1}{A(t_i)}$. Initial film height uncertainties were, on average, $\pm 5\%$. The dynamics of the film is recorded from $t=t_i$ onward. The 
angle of inclination was evaluated with image analysis and was evaluated within $\pm 0.7$ degrees.

In a typical experiment, the downstream contact line advances in the direction of the flow (Fig.~\ref{fig1}(d)). In the meantime, the film drains downstream and, in some cases, droplets were found to form and drip (Fig.~\ref{fig1}(b)). However, for greater inclinations no droplets were observed to form (Fig.~\ref{fig1}(c)). Describing the full dynamics requires complex non-linear and time evolving tools~\cite{kalliadasis2011falling} and is beyond the scope of this article. Instead, we propose to simply count the number of droplets formed during an experiment, keeping track of their dripping time and location. In particular, we focus on the central part of the film and denote $N$ the number of droplets dripping from this area. The region of interest, of area $A_i$, is defined by scaling the perimeter inwards of a length $2\pi\sqrt 2 \ell_c$ to avoid further complications at the contact line (see Supplementary Information for more details). 
To summarize, any experiment is fully characterized by the triplet: 
\begin{equation}
\label{ }
(h_i/\ell_c,\alpha,\bar N)_\text{exp}
\end{equation} 
where $\bar N= N/A_i$ denotes the density of droplets in the center part of the flow. Those data points are reported in Fig.~\ref{res} and are discussed in the next section. 

As a side remark, note that the ratio $h_i/\ell_c$ directly relates to the contact angle of the fluid on the substrate. In fact, the thickness of a partially wetting puddle sitting on an horizontal surface may be expressed as  $h_p=2\ell_c \sin{\theta_c/2} $, in the limit of very long times~\cite{de2002gouttes}. By assuming the initial film thickness being the one of the puddle, $h_p$, the results may be represented as a function of $\theta_c$ instead of $h_i/\ell_c$.


\section{Discussion}

As inferred from Fig.~\ref{exp}(a), the experiments divide into two categories:  experiments that do \textit{not} lead to any drop $\bar N=0$, and experiments during which several drops are formed $\bar N>0$. As anticipated, the largest drop-density experiments are the one with the smallest inclination of the substrate ($\alpha\simeq\pi/2$). In those cases, the flow is weak and does not affect the RTI. Increasing the inclinations (decreasing $\alpha$ when keeping other parameters fixed) yields a decrease in the number of drops. Interestingly, experiments reveal the existence of a critical angle $\alpha_c^\text{exp}$ under which no droplets are found. As seen from Fig.~\ref{exp}, this angle is a decreasing function of $h_i/\ell_c$. Those observations are now rationalized using the previous theoretical results. The absolute/convective transition angle $\alpha^*(h_i/\ell_c)$ from eq.~(\ref{criteria}) is superimposed to the experimental results in Fig.~\ref{exp}(a) and Fig.~\ref{exp}(b) in the form of a curved transparent surface and solid black line respectively. This theoretical prediction acts as a demarcation line between the two experimental regimes ($\bar N>0$ and  $\bar N=0$) over the range of experimental parameters explored.

 On the one hand, this agreement was expected as the physical mechanism behind a convective instability is precisely that the flow outruns the instability, which cannot grow locally, and therefore drop formation is prevented.  The geometry of the substrate, that is its inclination, stabilizes the flow. This type of geometrically induced stabilization is consistent with recent work on the dynamics of thin films on the underside of a cylinder where the curvature of the substrate, that is the local change in its slope, is shown to suppress the RTI if the film is thin enough~\cite{trinh2014curvature}.

On the other hand, we note the favorable agreement of theory with
the experiments despite the significant differences between the experimentally obtained flow and the model depicted in Fig.~\ref{fig2}. In particular, the model is one dimensional and our analysis is restricted to describing the linear behavior of the system. In reality, dripping is inherently non-linear and the experiment is essentially three dimensional. Yet, as it often turns out in fluid mechanics \citep{huerre1990local,huerre2000open}, the linear predictions prove to be an acceptable representation of the problem far beyond their validity range, thereby helping in the rationalization of the observed  physical reality.


\bibliography{RTinclinedplanePT}

\end{document}